# Design of the Inspection Process Using the GitHub Flow in Project Based Learning for Software Engineering and Its Practice

Yutsuki Miyashita
Graduate School of Education
Tokyo Gakugei University
Tokyo, JAPAN
m183310k @st.u-gakugei.ac.jp

Yuki Yamada
Graduate School of Education
Tokyo Gakugei University
Tokyo, JAPAN
m198132f @st.u-gakugei.ac.jp

Hiroaki Hashiura
Faculty of Advanced Engineering
Nippon Institute of Technology
Saitama, JAPAN
hashiura@nit.ac.jp

Atsuo Hazeyama
Department of Information Science
Tokyo Gakugei University
Tokyo, JAPAN
hazeyama@u-gakugei.ac.jp

## ABSTRACT

Project based learning (PBL) for software development (we call it software development PBL) has garnered attention as a practical educational method. A number of studies have reported on the introduction of social coding tools such as GitHub, in software development PBL. In education, it is important to give feedback (advice, error corrections, and so on) to learners, especially in software development PBL because almost all learners tackle practical software development from the viewpoint of technical and managerial aspects for the first time. This study regards inspection that is conducted in general software development activities as an opportunity to provide feedback and proposes the inspection process using the pull request on GitHub. By applying the proposed process to an actual software development PBL, we enable giving feedback to the accurate locations of artifacts the learners created.

## CCS CONCEPTS

## KEYWORDS

Software development PBL, Inspection, Feedback, GitHub flow, Pull request

## 1 Introduction

Software development nowadays is generally carried out as a project involving many members. Software development in the form of projects has been conducted in academic institutions as a practical educational method [8] (we call this practice software development PBL). Requirements analysis, design, implementation, and testing are done in software development. In each phase, artifacts such as requirements specification, class diagram, and so on are created. To create such artifacts, communication among stakeholders is important, and it is also important to manage the contents of communication.

Software engineering environments play an important role in managing artifacts and/or contents of communication. This applies to software development PBL. GitHub [13] has been used as a software engineering environment in open source software development. The use of GitHub in academic institutions is significant because they shoulder responsibility to bring up talented people in the industry [3]. Several studies have reported the use of GitHub in their practices [1, 2, 3, 7, 10]. However, the utilization of GitHub has not been presented clearly.

In education, it is important to give feedback (advice, error corrections, and so on) to learners, especially in software development PBL because most first time learners tackle practical software development from the viewpoint of technical and managerial aspects. This study takes into account the inspection [6], that is conducted in general software development activities, as an opportunity to provide feedback and proposes the inspection process on GitHub.

The remainder of this paper is organized as follows. Section II describes related work regarding this study. Section III provides an overview of our software development PBL. Section IV describes a brief introduction to GitHub flow. Section V proposes our inspection process on GitHub flow. Section VI presents a practice of the proposed inspection process. Section VII discusses the effectiveness of our proposal. Finally, Section VIII provides some concluding remarks.

## 2 Related work

This section introduces studies that have used GitHub in software development PBL.

Francese et al. report their experience in developing a smart phone application in the form of a project using GitHub [2]. The development process they adopted is incremental prototyping. They asked the students to create a project proposal, requirements analysis document, some kinds of diagrams (use case diagram, class diagram, and so on), user interface mockup, and specification of black box testing. They use the milestones, issue and label features provided by GitHub. They also claim that the transparency of GitHub is important as it helps in ascertaining who does what, thus enabling students to learn from the actions of each other.

Haaranen and Lehtinen study how GitHub should be taught to students [3]. They do not target project-based software development.

Zakiah and Fauzan propose a collaborative software development learning model using GitHub [10]. They divide learners into two roles, leaders, and members. They present models on what each role does in the Git flow.

Feliciano et al. demonstrate the importance of GitHub from the viewpoint of students by an experiment [1]. They show the benefits of viewing the activities of other members, including history, contribution to the activities of others, learning technologies used in the industry in a practical manner, and interacting with people who are external to the project using GitHub. In addition, they raise some issues such as the outcomes becoming public, unfamiliarity with GitHub flow, information overload, and so on.

Raibulet and Francesca report using GitHub in a software engineering course for third year undergraduate students [7]. The students were taught the functions of GitHub (branch, push, pull, etc.) via lectures. They were asked to submit an executable system via analysis, design, and implementation of a presented task during a one-month project. They used GitHub as a platform for collaborative endeavor.

These studies focus on the benefits of using GitHub in software engineering education, the functions of GitHub used during the practice, and the teaching of GitHub. However, these studies do not address the design of the effective process of using GitHub for software development PBL.

## 3 Software Development PBL at Tokyo Gakugei University

This section presents an overview of our software development PBL.

### 3.1 Lecture before software development PBL

Our software development PBL is introduced as an elective subject for third year undergraduate students who major in informatics education (full quota of fifteen students). We offer a subject called "design of information systems," just before the PBL that introduces the basics of software engineering. This subject also deals with simple web application development using JSP/Servlet technologies and relational database management systems.

### 3.2 Software development PBL

Here, we briefly introduce our software development PBL (details can be found in [4]). Our software development PBL organizes students into groups which are consisted of three to five members. The number of groups is two or three, therefore, our PBL is small scale. Groups are organized by the instructor, based on responses to a questionnaire regarding participation in group work and the grade of "design of information systems." The development process follows the waterfall model. First, each group conducts requirements analysis from several lines of requirements sentences given by the instructor, and creates the requirements specification. Following requirements analysis comes the upstream process, in which each group creates artifacts such as a user interface (UI) design document, a class diagram, a database (DB) design document, sequence diagrams, and a state chart. Each group asks the instructor and the teaching assistant (TA) to inspect the artifacts (we call the instructor and TA together as the teaching staff). Master course students who have cleared this PBL take charge as TAs. The groups are asked to assign their members to create sequence diagrams and source codes, and doing unit testing per use case defined in the requirements specification. After unit testing, the developed functions are integrated into a system and uploaded to a server. Primary sequences are checked, and then the group releases the system. The group performs system testing and the teaching staff performs acceptance testing in parallel.

The groups are required to revise the requirements specification whenever the specification changes, because inspection and acceptance testing are done based on the requirements specification.

Some major milestones (due date for inspection request, release for acceptance testing, and project completion) are presented to the groups by the instructor. Other schedules, except for the schedules determined by the instructor, can be determined by the groups because of their acquisition of project management skills.

### 3.3 Usage of GitHub in Software Development PBL

*3.3.1 Artifact management.* Our PBL manages all artifacts using Git. Artifacts created in the upstream process such as the requirements specification, are required for description in the markdown notation. Unified Modelling Language (UML) diagrams are drawn using an open source text-based language called PlantUML [14]. Managing all artifacts in Git enables software development along the development flow called GitHub flow in the upstream process as well as in the downstream process. We introduce GitHub flow in section 4.

*3.3.2 Project management.* We use the milestone function in GitHub for project management. The milestone function of GitHub enables task management by setting goals (deadlines) to artifacts and attaching issues and/or pull requests to milestones.

Figure 1 shows an example of the milestone function. It depicts the creation of requirements specification, UI design, class diagram, and DB design, their schedules and the status of their progress.

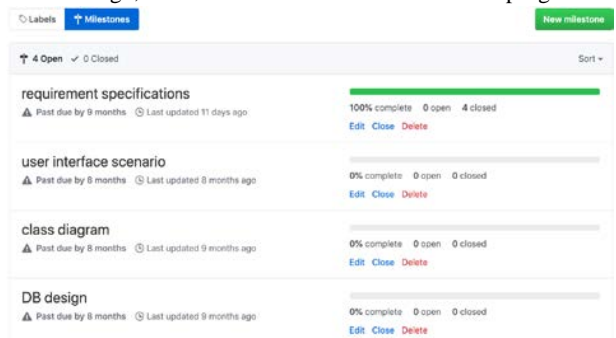

**Figure 1: Example of Milestones.**

### 3.4 Inspection Process

The groups request the teaching staff to perform inspection, once they have prepared their artifacts. Artifact reading is performed by the instructor and the TA individually. When they all finish reading the artifacts and giving comments, the instructor checks them and then notifies the results to the groups. Groups can request inspection for each artifact twice at most. When groups request inspection for the second time, revised artifacts as well as replies to the comments given in the first inspection are asked to be submitted. With this information, the teaching staff can then ascertain the intention of the revision.

The 90-minutes class is held once a week. The groups and the teaching staff meet during this period. The duration of the project is fourteen weeks, and therefore, a face-to-face inspection for everyone in the classroom is not realistic. Therefore, we do the inspection process online in the development environment in an asynchronous way. In addition, the inspection process is followed up in the classroom if necessary.

We have developed an environment for our PBL and have successfully run it [5]. We have used GitBucket [12] from 2015 to 2017. We have been using GitHub [13] since 2017.

## 4 GitHub Flow

GitHub flow is a lightweight branch-based workflow [11]. The master branch is always enabled and deployment ready. Developers make a new branch in implementing and/or revising a function. When merging into the master branch, developers create a pull request and ask members for reviews.

## 5 GitHub Flow

Our proposed inspection process asks for inspection using the pull request. We describe the requirements for inspection in GitHub flow.

- Comments are given on each line that comprises an artifact. This enables inspectors to give comments, without specifying the locations where comments are given. In other words, the inspectors can focus on the contents and descriptions of the feedback.
- Review comments are given to differentiate between the current and the previous versions at the time of the second inspection, so that burden of the inspectors decreases.
- The teaching staff is notified of inspection requests.

We ensure that artifacts are created in text format as much as possible, so that we can use the 'diff' function Git provides in the markdown notation or PlantUML. Previously, many types of artifacts were created using word processing or presentation software. Therefore, inspection could not be done in a pull request base but was done in an issue base.

Unlike industrial software development, in our software development PBL, not only developers but also the teaching staff participate in the inspection process. We require the groups to review among their members, before requesting for inspection by the teaching staff.

In the 2018 PBL, we found that parts of the artifacts were not shown as differences at the time of inspection; because they were merged into the master branch before inspection was done. Once an artifact is merged into the master branch, differences do not appear. Therefore, it is impossible to give comments line by line. We should avoid the artifact being merged into the master branch after the group review, until the inspection is finished.

Figure 2 shows the proposed inspection flow. First, a group creates a branch for inspection from the master branch and then the learners create branches for each individual work from the created inspection branches to create documents. Learners create the pull request between the inspection branch and their work branch, and perform reviews within the groups. Learners merge their work branch into the inspection branch by creating the pull request. This attaches labels of inspection, and designates the teaching staff for inspection. The inspection request is thus completed.

Figure 3 is a screen shot of inspection requests from learners. The designated teaching staff can do the inspection. Figure 4 shows an example screen shot that of the teaching staff writing inspection comments. If the second inspection is required, a pull request is created by creating a branch for the second inspection that merges to the branch for the first inspection. The learners thus ask the teaching staff for the second inspection. This enables to review the differences between the artifact created at the first inspection and the second. After inspection, learners commit revision, and merge the revision into the master branch.

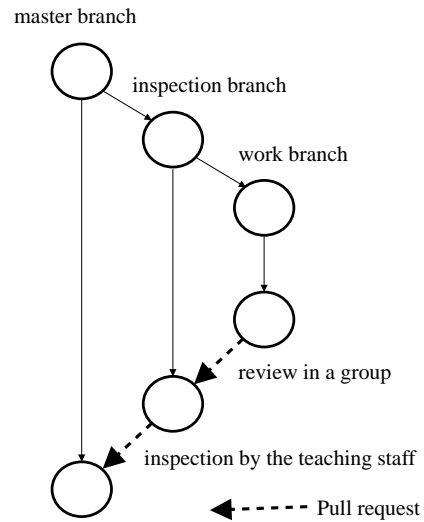

**Figure 2: Proposed inspection flow.**

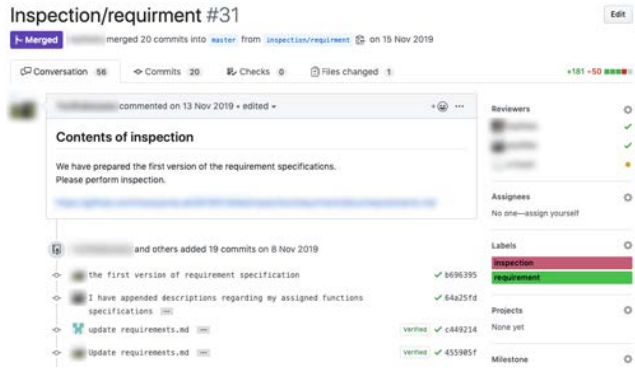

**Figure 3: A screen shot a learner requests for inspection.**

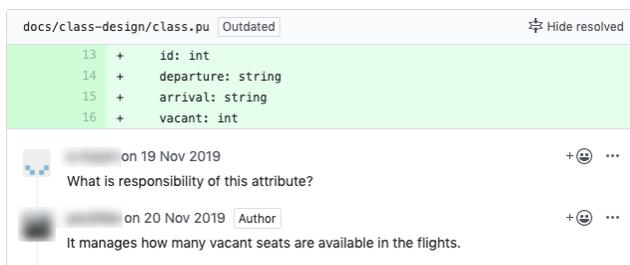

**Figure 4: A screen shot the teaching staff write inspection comments.**

## 6 Practice

This section describes the practice of the proposed inspection process using GitHub flow in the 2019 software development PBL.

Nine students took this PBL, and were organized into two groups of four and five members. As Figure 2 shows, we confirm that both groups perform group review before submitting the pull requests for inspection to the teaching staff. Figure 5 is an example screenshot of the pull request in a group. "create_db_design" is a work branch. This group creates the pull request when it merges with the inspection branch "inspection/db-design."

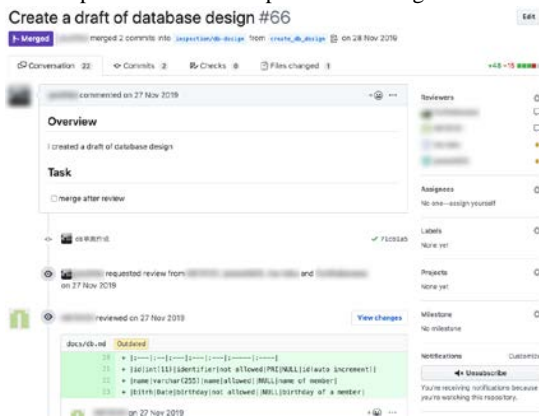

**Figure 5: A screen shot of review within group.**

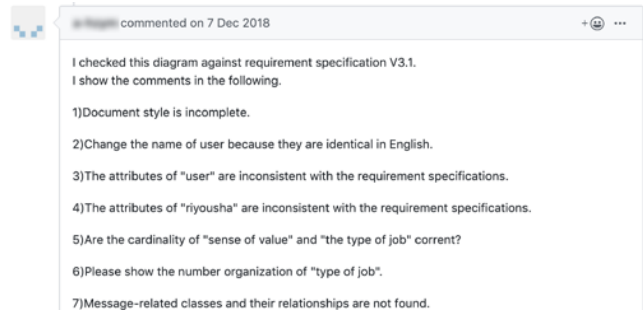

**Figure 6: Example of class diagram inspection in 2018.**

## 7 Discussions

This section discusses the three viewpoints we raised as the requirements for the inspection flow in section V.

### 7.1 To give comments to each line of an artifact

Figure 6 is a screenshot depicting inspection using the issue in the 2018 software development PBL. In this instance, all comments are written in one issue. On the other hand, in the proposed inspection process, as shown in Figure 7, a class diagram is created by using PlantUML, and inspection is initiated via the pull request. By inspecting using the pull request to the text-based artifact, the inspector can directly give comments at the location of the problem. Furthermore, as shown in Figure. 7, inspectors and developers can discuss each comment at the location.

### 7.2 To give review comments to the differences between the current version and the previous one

In our experience, several comments are given to requirements specification in general. Therefore, the volume of modifications is large. We asked the groups to create the pull request at the second inspection from the first inspection branch. Thus the inspectors can ascertain where the groups add, modify, or delete the document according to the first inspection comments. All groups followed this instruction and the inspectors could ascertain the differences as expected. On the other hand, as the volume of modifications in the class diagram is not so large as compared to the requirements specification, the TA announces that the modification commit is made in the first pull request, and learners notify the teaching staff of the completion of their modification. However, we found that it is difficult for the teaching staff to confirm how the modification is made. For example, as Figure 8 shows, although a group presents a modification commit in a response, it is difficult to understand the modifications from the name and volume (the value is 0). We find that it is important to give guidance so that learners can create appropriate commits and write clear commit messages.

### 7.3 To notify the teaching staff of request for inspection

Design of the Inspection Process Using the GitHub Flow in
Project Based Learning for Software Engineering and Its Practice

We gave a lecture on the proposed inspection process including setting the teaching staff to be reviewers at the pull request, and attaching inspection labels before starting the PBL. The teaching staff could differentiate between pull requests within groups, and in inspections. Although it is not easy for learners to understand the GitHub flow [1, 9], we think effective feedbacks (direct comments and discussions at the location where problems occur) are given by creating artifacts in the form of markdowns and PlantUML, and performing inspections on them.

## 8 Conclusions

We regard software inspection as an opportunity for feedback, which is important in education. We have proposed the inspection process on GitHub flow, to run it on GitHub. We have reported a practice of this process in software development PBL. The proposed inspection process has enabled the teaching staff to give comments to the artifacts created by the learners, line by line, and provide apt feedback. Although it is not easy for students to be familiarized with GitHub flow, the proposed process contributes to give apt feedback to the learners and to reduce the burden of inspection on the teaching staff.

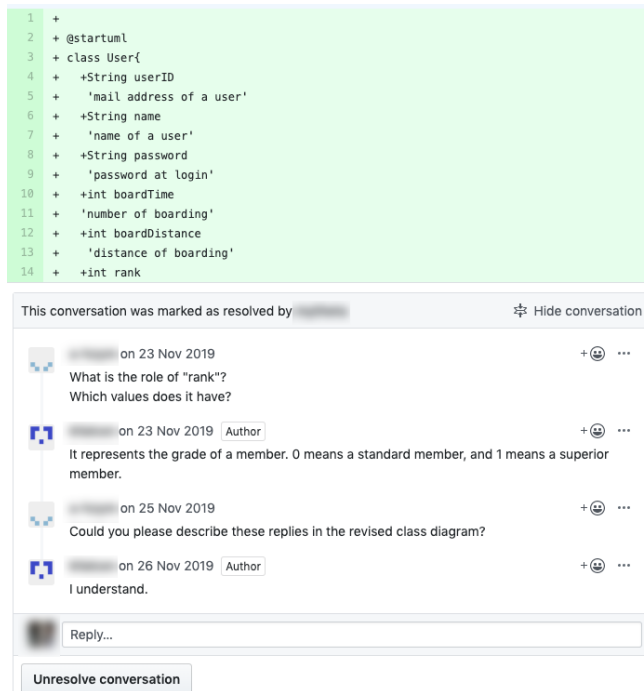

**Figure 7: Discussions between the group and the teaching staff for each comment in 2019.**

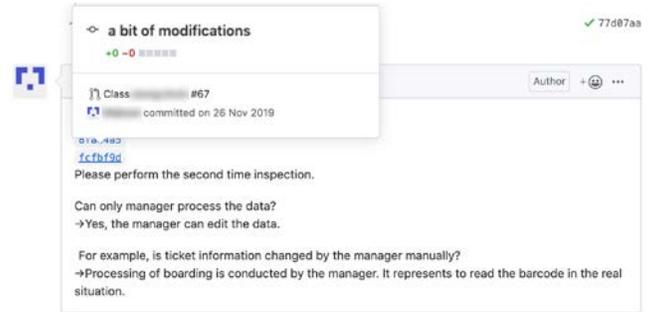

**Figure 8: Example of an inappropriate modification commit.**

## ACKNOWLEDGMENTS

This study is partially supported by the Grant-in Aid for No. (C) 18K11579 from the Ministry of Education, Science, Sports and Culture of Japan.

## REFERENCES

[1] Joseph Feliciano, Margaret-Anne Storey, and Alexey Zagalsky, Student experiences using GitHub in software engineering courses: a case study, Proceedings of the 38th International Conference on Software Engineering Companion, pp. 422-431, ACM, 2016.
[2] Rita Francese, Carmine Gravino, Michele Risi, Giuseppe Scanniello, and Genoveffa Tortora, On the Experience of Using GitHub in the Context of an Academic Course for the Development of Apps for Smart Devices, DMS, pp. 292-299, 2015.
[3] Lassi Haaranen and Teemu Lehtinen, Teaching Git on the Side: Version Control System as a Course Platform, Proceedings of the 2015 ACM Conference on Innovation and Technology in Computer Science Education, pp. 87-92, ACM, 2015.
[4] Atsuo Hazeyama, A Case Study of Undergraduate Group-based Software Engineering Project Course for Real World Application, Proceedings of the First International Symposium on Tangible Software Engineering Education, pp. 39-44, 2009.
[5] Atsuo Hazeyama, Collaborative Software Engineering Learning Environment Associating Artifacts Management with Communication Support, Proceedings of the 2014 IIAI 3rd International Conference on Advanced Applied Informatics (IIAIAAI), IEEE Computer Society, pp.592-596, 2014.
[6] Sami Kollanus, and Jussi Koskinen, Survey of Software Inspection Research, The Open Software Engineering Journal, 3(1), 15-34, 2009.
[7] Claudia Raibulet and Fontana Arcelli Francesca, Collaborative and Teamwork Software Development in an Undergraduate Software Engineering Course, Journal of Systems and Software, 144, pp. 409-422, 2018.
[8] Till Schümmer, Stephan Lukosch, and Joerg M. Haake, Teaching Distributed Software Development with the Project Method, Proceedings of the 2005 Conference on Computer Support for Collaborative Learning: Learning 2005: the Next 10 Years!, pp. 577-586, International Society of the Learning Sciences, 2005.
[9] Haruaki Tamada, Learning support system of GitHub flow for novice developers, Proceedings of the Software Engineering Symposium 2019 (SES2019) Workshop, 2 pages, IPSJ, 2019 (In Japanese).
[10] Azizah Zakiah and Mohamed Nurkamal Fauzan, Collaborative Learning Model of Software Engineering Using Github for Informatics Student, Proceedings of the 4th International Conference on Cyber and IT Service Management, pp. 1-5, IEEE, 2016.
[11] GitHub flow, https://guides.github.com/introduction/flow/ (accessed 5 November 2019).
[12] GitBucket, https://gitbucket.github.io/ (accessed 5 November 2019).
[13] GitHub, https://github.com/ (accessed 5 November 2019).
[14] Plant UML, https://plantuml.com/ (accessed 5 November 2019).